\documentclass[pra, aps, twocolumn, groupedaddress, showpacs, superscriptaddress]{revtex4-2}
\usepackage{amssymb, amsmath, amsthm, color, graphicx, times, graphicx}
\usepackage{hyperref}
\usepackage{subfigure}
\usepackage{times}
\usepackage{bbold}
\usepackage{color}

\providecommand{\openone}{\leavevmode\hbox{\small1\kern-4.3pt\normalsize1}}

\theoremstyle{plain}

\theoremstyle{definition}

\usepackage{orcidlink}
\makeatletter
\newsavebox{\@brx}
\newcommand{\llangle}[1][]{\savebox{\@brx}{\(\m@th{#1\langle}\)}%
  \mathopen{\copy\@brx\mkern2mu\kern-0.9\wd\@brx\usebox{\@brx}}}
\newcommand{\rrangle}[1][]{\savebox{\@brx}{\(\m@th{#1\rangle}\)}%
  \mathclose{\copy\@brx\mkern2mu\kern-0.9\wd\@brx\usebox{\@brx}}}
\makeatother

\begin{document}
\title{Quantum phase estimation and realistic detection schemes in Mach-Zehnder interferometer using SU(2) coherent states}
\author{M. Abdellaoui\orcidlink{0009-0002-0801-6042}}\affiliation{LPHE-Modeling and Simulation, Faculty of Sciences, Mohammed V University in Rabat, Rabat, Morocco.}
\author{N.-E. Abouelkhir \orcidlink{0000-0002-6164-0525}}\affiliation{LPHE-Modeling and Simulation, Faculty of Sciences, Mohammed V University in Rabat, Rabat, Morocco.}
\author{A. Slaoui \orcidlink{0000-0002-5284-3240}}\email{Corresponding author: abdallah.slaoui@um5s.net.ma}\affiliation{LPHE-Modeling and Simulation, Faculty of Sciences, Mohammed V University in Rabat, Rabat, Morocco.}\affiliation{Centre of Physics and Mathematics, CPM, Faculty of Sciences, Mohammed V University in Rabat, Rabat, Morocco.}
\author{R. Ahl Laamara \orcidlink{0000-0001-8410-9983}}\affiliation{LPHE-Modeling and Simulation, Faculty of Sciences, Mohammed V University in Rabat, Rabat, Morocco.}\affiliation{Centre of Physics and Mathematics, CPM, Faculty of Sciences, Mohammed V University in Rabat, Rabat, Morocco.}

\begin{abstract}
\textbf{Abstract}: In quantum parameter estimation, the quantum Cramér-Rao bound (QCRB) sets a fundamental limit on the precision achievable with unbiased estimators. It relates the uncertainty in estimating a parameter to the inverse of the quantum Fisher information (QFI). Both QCRB and QFI are valuable tools for analyzing interferometric phase sensitivity. This paper compares the single-parameter and two-parameter QFI for a Mach-Zehnder interferometer (MZI) with three detection schemes: single-mode and difference intensity detection, neither has access to an external phase reference and balanced homodyne detection with access to an external phase reference. We use a spin-coherent state associated with the su(2) algebra as the input state in all scenarios and show that all three schemes can achieve the QCRB for the spin-coherent input state. Furthermore, we explore the utilization of SU(2) coherent states in diverse scenarios. Significantly, we find that the best pressure is obtained when the total angular momentum quantum number $j$ is high, and we demonstrate that given optimal conditions, all detection schemes can achieve the QCRB by utilizing SU(2) coherent states as input states.

\par
\vspace{0.25cm}
\textbf{Keywords:} Mach-Zehnder interferometer, Quantum Fisher information, Quantum Cramér-Rao bound, SU(2) coherent states, Phase sensitivity.
\end{abstract}
\date{\today}

\maketitle
\section{Introduction}
Fundamentally, the laws of quantum mechanics govern all processes of measurement. Quantum mechanics directly influences the precision of measurements that can be performed by imposing constraints. These constraints directly influence various areas of physics and technology \cite{Peters2001, Abbott2017}. 
However, they also enable the development of innovative measurement approaches with improved performance\cite{Abbott2016, Slaoui2023}, leveraging non-classical phenomena \cite {Taylor 2016}. Quantum metrology involves studying the impact of quantum mechanics on measurement systems and the development of new measurement technologies based on non-classical effects\cite{Birrittella2012, Gerry2012, Riedel2010}. Generally, quantum metrology protocols comprise three primary steps: Firstly, the probe is prepared in an initial state; secondly, it undergoes evolution under the influence of the quantum process described by a physical parameter; and finally, the encoded state is measured. This state is calculated by using appropriate measurement settings, and the measurement results are then processed to estimate the considered parameter.\par

Precision measurement is a critical element in science as well as in technology\cite{Liu2010, Appel2009}. Indeed, advancements in measurement techniques have led to many important discoveries. More sensitive instruments, such as telescopes and microscopes, have been instrumental in discovering and comprehending new phenomena to verify or invalidate theoretical predictions. Improving measurement sensitivity is a cardinal element in advancing science and technology\cite{Ataman2018}. Interferometry, especially the Mach-Zehnder interferometer (MZI)\cite{Xiao1987, Pezze2013, Anisimov2010}, is an exceedingly delicate and widely used measurement technique\cite {Stefan2022}. The phase sensitivity of interferometry is an important research topic in several rapidly growing scientific fields, including gravitational wave detectors and quantum technology\cite{Demkowicz2013, Grote2013, Oelker2014, Tse2019}.\par

Improving interferometry's phase sensitivity has both a classical and a quantum purpose. Hence, quantum metrology has provided the opportunity to surpass the classical shot-noise limit into the so-called quantum or sub-shot-noise regime\cite{Ataman2022, Vahlbruch2018, Xiao1987}. By using classical resources at the input of an interferometer, one can reach the shot-noise limit (SNL), also known as the standard quantum limit (SQL), for phase sensitivity. The SQL is given by $\Delta\varphi_{SQL}\sim 1/\sqrt{N}$, where $N$ is the average number of input photons \cite{Caves1981, Giovannetti2011}. It has long been shown that the shot-noise-limited single coherent input interferometer can be surpassed by non-classical light states\cite{Ataman2018}. The coherent plus squeezed vacuum input state has gained popularity due to its favorable performance in both low- and high-power regimes\cite{Mishra2022}. In a recent study, the squeezing technique was demonstrated to have two potential applications: reducing fluctuations in laser power and detecting motion in mechanical oscillators. By applying squeezed, NOON\cite{Berry2009, Gerry2010, Dowling2008}, or other non-classical input states at the input of an interferometer, i.e., by using quantum resources \cite{Ou1997}, one can approach the ultimate quantum limit $\Delta\varphi_{HL}\sim 1/N$, known as the Heisenberg limit (HL) \cite{Gerry2001, Gerry2002}.\par

Sparaciari et al. \cite{Sparaciari 2015, Sparaciari 2016} provide a solid theoretical foundation for understanding the limits of phase estimation using Gaussian states, focusing on optimal configurations and the role of squeezing under idealized conditions. this paper builds on this theoretical foundation by addressing practical aspects of detection and noise, making significant contributions toward the experimental realization of high-precision phase estimation in Mach-Zehnder interferometers. While the works of Sparaciari et al. are instrumental in setting the theoretical limits and identifying optimal state configurations using Gaussian states, this manuscript adds an essential layer of practicality. It focuses on SU(2) coherent states of spin. This approach ensures that the theoretical insights can be effectively translated into experimental practice, marking a significant advancement in quantum interferometry.\par 

In this paper, we focus on the phase sensitivity of an MZI consisting of two beamsplitters\cite{Mishra2022}. Most interferometers can be converted to MZI to optimize phase sensitivity for various input states and detection schemes. The quantum Fisher information (QFI) and the associated quantum Cramér-Rao bound (QCRB) are powerful tools \cite {Chang 2020, Abouelkhir2023, Braunstein1994, Mohamed2022 EPJ, Abdel-Aty2023} for setting upper-performance bounds in phase estimation\cite {Mohamed2022}. The phase sensitivity of an MZI depends on several factors \cite{Mishra2022}, including the input state and the detection scheme chosen.\par

In quantum physics, coherent states are essential in encoding quantum information on continuous variables\cite{Ikken2023}, particularly in quantum optics. Perelomov introduced spin coherent states (SCS) as a typical type of cohesive state in su(2) Lie algebras\cite {Berrada2019}. The similarity of the Lie algebras corresponding to su(1,1) and su(2), respectively, implies a close relationship between the SU(1,1) and SU(2) coherent states\cite {Ou2020}. This paper uses a cohesive state in the input called the spin-coherent states associated with the su(2) algebra\cite{Yurke1986}. 

This paper follows a structured organization. Section II briefly recalls the SU(2) coherent states. Moving forward to section III, We'll introduce specific conventions and present the notion of a two-parameter QFI. Then, we approach the single-parameter QFI, considering both asymmetric and symmetric phase shift scenarios. Section IV provides expressions for the QFIs in all three considered scenarios for input SU(2) coherent states. The three detection schemes considered are described in section V. Moving to section VI, we detail and discuss the performance of these detection schemes with input SU(2) coherent states. Finally, section VII summarises our work and concludes this study.

\section{Algebraic Foundation and Coherent States in the SU(2) Interferometer for Precise Phase Estimation}

The algebra responsible for the integer spin representation is the special Lie algebra, also known as the Lie algebra of rotations in space\cite{Benatti2011}. This algebra is denoted $su(2)$ and is associated with the Lie group SU(2). It is generated by three generators, corresponding to the three components of angular momentum or spin. These generators are traditionally denoted $Sx$, $Sy$, and $Sz$. The Lie algebra $su(2)$ has the following commutation relations
\begin{align} \nonumber
&\left[\hat{S}_{+}, \hat{S}_{-}\right]=\hat{S}_z,
&\left [\hat{S}_{\pm}, \hat{S}_{z}\right]=\mp\hat{S}_{x},
\end{align}
Where the operators ${S}_{+}$, ${S}_{-}$, and ${S}_{z}$ are Hermitian, meaning they are equal to their adjoint (transposed conjugate)
\begin{equation}
 \hat S_+^\dagger = \hat S_+, \quad \hat S_-^\dagger = \hat S_-, \quad \hat S_z^\dagger = \hat S_z.
\end{equation}

The Lie algebra $su(2)$ is fundamental in quantum mechanics for describing the spin behavior of particles. The eigenvalues of the $S_+$, $S_-$, and $Sz$ operators allow us to determine the possible values of the spin in a given quantum system.

The spin-coherent states are described as a single-mode collection. Then, the HPR version of the spin Lie algebra is related to Bose annihilation and creation operators \cite{Berrada}. These operators are represented as follows
\begin{equation}
S_+ = \hat{b}^\dagger \sqrt{2s - \hat{n}}, \quad S_- = \sqrt{2s - \hat{n}} \hat{b}, \quad \hat{S}_z = \hat{n} - s
\end{equation}

In addition, assuming that $a^\dagger$ and $a$ conform to the Bose algebra $[b b^\dagger] = 1$  where the symbol $s$ represents the total angular momentum quantum number, in this context, the angular momentum states $ \lvert s, m \rangle$ are interconnected with the Bose number states $\lvert n \rangle$. This connection is expressed as $ \lvert s, m \rangle \sim \lvert n \rangle$, where $ n = s + m$. It is worth noting that only the numerical states corresponding to $n = 0, 1, ...,2s$ contribute to constructing representations for spin algebra with a fixed value of $s$.
Currently, our objective is to identify the input state that enables the most precise phase estimation, specifically achieving the highest Quantum Fisher Information (QFI). \\
The input state we are interested in is a pure two-mode one, a superposition of macroscopic spin-coherent states. This state can be conceptualized as a combination of NOON states\cite{Berrada2012}, which are frequently employed in quantum metrology due to their remarkable precision enhancement capabilities. Then, in mode $i$, the spin-coherent state could be described with the number of states using the Holstein-Primakoff realization (HPR) as follows
\begin{equation} \label{input state}
 |\lambda, j\rangle=\frac{1}{\sqrt{{ }_1 F_0^{(-)}\left(2 j,|\lambda|^2\right)}} \sum_{\eta=0}^{2 j}\left[\frac{(2 j)_{-\eta}}{\eta!}\right]^{\frac{1}{2}} \lambda^\eta|\eta\rangle
 \end{equation}
Note that $(\Lambda)_{-\eta} $ is a negative Pochammer symbol with $(\Lambda)_0 = 1$ such
\begin{equation}
(\Lambda)_{-\eta} = \Lambda(\Lambda - 1) \cdots (\Lambda - \eta + 1)
\end{equation}
So we find that
\begin{equation}
\begin{aligned}
(2 j)_{-\eta} & =\underbrace{2 j(2 j-1) \cdots(2 j-\eta+1)} \\
& =\frac{2 j !}{(2 j-\eta) !}
\end{aligned} 
\end{equation}
We also have the expression of the negative hypergeometric function by
\begin{equation}
{ }_1 F_0^{(-)}(P, x)= \sum_{\eta=0}^P(P)_{-\eta} x^\eta / \eta!
\end{equation}
By using this definition, we can arrive at the following equation
\begin{equation}
{ }_1 F_0^{(-)}\left(2 j,|\lambda|^2\right)=\sum_{\eta=0}^{2 j}(2 j)_{-\eta}|\lambda|^{2 \eta} / \eta !
\end{equation}
Using the definition mentioned in \cite{Berrada2012} and the preceding equation, we will obtain
\begin{equation}
\begin{aligned}
{ }_1 F_0^{(-)}\left(2 j,|\lambda|^2\right) & =\sum_{m=0}^{2 j}(2 j)_{-m}|\lambda|^{2 m} / m ! \\
& =\sum_{m=0}^{2 j} \frac{2 j !}{(2 j-m) !}|\lambda|^{2 m} / m !
\end{aligned}
\end{equation}
So the expression for our spin-coherent state $|\lambda, j\rangle_i$ in mode $i$ will be written as follows
\begin{equation}
|\lambda, j\rangle =\frac{1}{\sqrt{\sum_{m=0}^{2 j} \frac{2 j !}{(2 j-m) !}|\lambda|^{2 m} / m !}} \sum_{\eta=0}^{2 j}\left[\frac{(2 j)_{-\eta}}{\eta !}\right]^{\frac{1}{2}} \lambda^\eta|\eta\rangle
\end{equation}
This will lead us to the following expression
\begin{equation}
|\lambda, j\rangle=C(|\lambda|) \sum_{\eta=0}^{2 j} \frac{1}{\sqrt{\eta !(2 j-\eta) !}} \lambda^\eta|\eta\rangle 
\end{equation}
Where
\begin{equation}
    C(|\lambda|)=\frac{1}{\sqrt{\sum_{m=0}^{2 j} \frac{1}{(2 j-m) !}|\lambda|^{2 m} / m !}}
\end{equation}
The parameter $\lambda$ is determined by 
$\lambda = e^{-i\phi} \tan\left(\frac{\theta}{2}\right)$, and can cover the entire complex plain with $0 \leq |\lambda| < \infty$.

In the SU(2) interferometer, the photon number operator is defined by $n_1=b_1^+b_1$. From this vantage point, we will discover the subsequent.
\begin{equation}
    \langle \hat{n}_1 \rangle = C(|\lambda|)^{2} \sum_{n=0}^{2j} \frac{n|\lambda|^{2n}}{n!(2j-1)!}
\end{equation}

and also for $n^2$
\begin{equation}
\langle \hat{n}_1^{2}\rangle=C(|\lambda|)^{2} \sum_{n=0}^{2 j} \frac{n^2|\lambda|^{2 n}}{n !(2 j-1) !}
\end{equation}

\section {Quantum Interferometry and Parameter Estimation in Mach-Zehnder Interferometers} \label{section 3}

\begin{figure}[ht] 
    \centering
		\includegraphics[width=\linewidth]{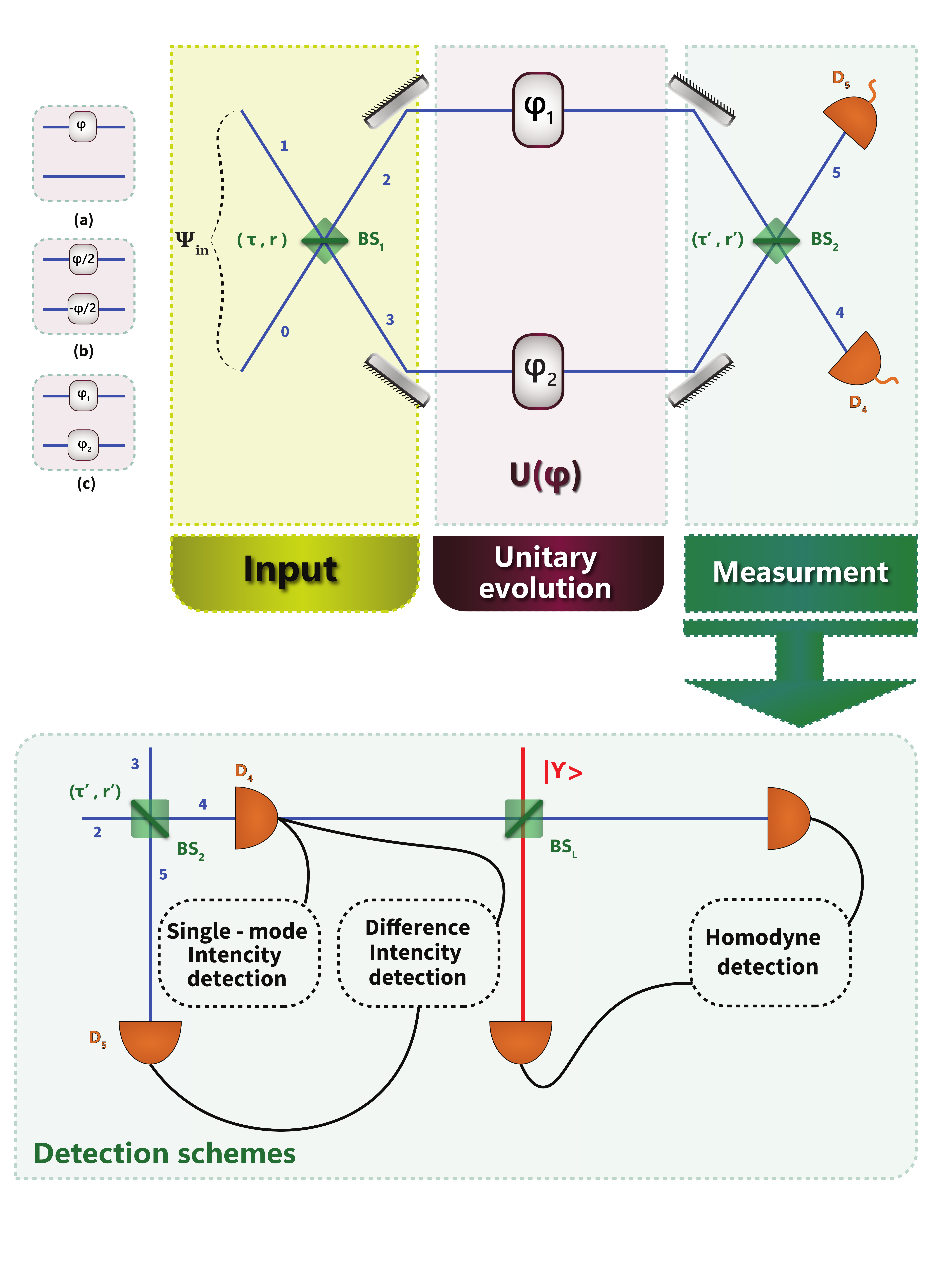}
	\caption{Exploration Realistic Detection Schemes in Mach-Zehnder (MZ) Interferometry: Differential Intensity, Single-Mode Intensity, and Balanced Homodyne Detection.} 
   \label{detection schemes}
\end{figure}

We are considering using the conventional Mach-Zehnder (MZ) interferometric configuration shown in Fig. 1, where the two beam splitters, BS1 and BS2, possess transmission (reflection) coefficients denoted as $\tau(r)$ and $\tau'(r')$, respectively. Throughout our study, we assume that the input state is pure and incurs no losses \cite{Pezze2008}. \\
Generally, the precision of phase estimation in quantum interferometry is constrained by QFI, which quantifies the ultimate limit of precision achievable in estimating the phase shift\cite{Lang2013, Lang2014}. This information metric depends on how the phase delay in the interferometer is modeled, reflecting the sensitivity of the interferometric setup to variations in the phase. This modeling includes (a) a single-phase shift in the lower arm; (b) two symmetrically distributed phase shifts, $\pm\varphi/2$; and (c) two independent phase shifts, $\varphi_{1} (\varphi_{2})$, in the upper (lower) arm. We begin by looking at the scenario of a phase shift in one arm that occurs at output 3 of BS1. We use the notations in Fig.\ref{detection schemes} to determine that $ |\Psi\rangle = e^{-i\varphi \hat{n}_3} |\psi_{23}\rangle $.

Using the Quantum Fisher Information Matrix (QFIM) definition for a pure parameterized state\cite{ Liu2017, Liu2020, Abouelkhir2023(2)}, where QFIM is represented as a $2x2$ matrix
\begin{equation} \label{QFIM}
    \mathcal{F}=\left(\begin{array}{ll}
\mathcal{F}_{dd} & \mathcal{F}_{sd} \\
\mathcal{F}_{ds} & \mathcal{F}_{ss}
\end{array}\right)
\end{equation}

The single-parameter QFI\cite{Ataman2022}, denoted as $\mathcal{F}^{(a)}$, is given by
\begin{equation}
 \mathcal{F}^{(a)} = 4\Delta^2 \hat{n}_3,
\end{equation}

The Cramer-Rao quantum limit (QCRB) sets the lower bound on the variance of any unbiased parameter estimator\cite{Liu2016, Helstrom1968}. Also, it provides the lowest value at which the parameter can be estimated. This quantum limit in single-parameter estimation can be expressed using the following formula \cite {Rao 1945, Cramer 1946}

\begin{equation} \label{Var}
Var(\hat{\varphi}) \geq \frac{1}{M}\mathcal{F}^{-1}
\end{equation}
Where $M$ symbolizes the number of repeated experiments, $Var$ is the estimator's variance matrix, and $\mathcal{F}$ is the QFIM \cite{Liu2020}.

The bound indicated in equation (\ref{Var}) implies that the inverse of the Fisher information limits the variation in any unbiased estimator for the parameter $\theta$. This suggests that an estimator's variance is at least proportional to the Fisher information's inverse. Thus, higher estimate precision corresponds with higher Fisher information. Thus, the Fisher information is vital in determining the maximum value for the accuracy obtained in a parameter estimate sit\cite{Lu2012}. In this context, the Fisher information captures the maximum amount of information that can be gathered about the true value sought for the parameter $\theta$. In our case, we found 
\begin{equation}
    \Delta\varphi_{QCRB}^{a} = \frac{1}{\sqrt{\mathcal{F}^{(a)}}}
\end{equation}
 
By utilizing the field operator transformations 
\begin{equation} \label{transformations1}
\begin{bmatrix}
\hat{b}_{2}\\
\hat{b}_{3}
\end{bmatrix}=\begin{pmatrix}
\tau & r\\
r & \tau
\end{pmatrix}\begin{bmatrix}
\hat{b}_{0}\\
\hat{b}_{1}
\end{bmatrix}
\end{equation}
We have the following equations: $\tau ^*r + \tau r^* = 0$ and $|r|^2 + |\tau|^2 = 1$.This relation implies that $\tau ^*r = \pm i\tau r$. This work will use the convention $\tau ^*r = i|\tau r|$ to preserve generality
we obtain
{\small
\begin{align} \label{F(b)}
\mathcal{F}^{(a)}=&4|r|^4\Delta^2\hat{n}_{0}+4\tau ^4\Delta^2\hat{n}_{1}\\
&+4|\tau r|^4\left(\langle\hat{n}_{0}\rangle+\langle\hat{n}_{1}\rangle+2\left(\langle\hat{n}_{0}\rangle\langle\hat{n}_{1}\rangle-|\langle\hat{b}_{0}\rangle|^2|\langle\hat{b}_{1}\rangle|^2\right)\right)\notag\\
&-8|\tau r|^2\mathfrak{r}\left\{\langle\hat{b}^2_{0}\rangle\langle(\hat{b}^{\dagger}_{1})^2\rangle-\langle\hat{b}_{0}\rangle\langle\hat{b}^{\dagger}_{1}\rangle\right\}-8|\tau r|\mathfrak{I}\left\{\langle\hat{b}_{0}\rangle\langle\hat{b}^{\dagger}_{1}\rangle\right\}\notag\\ \nonumber
&-16|\tau r||r|^2\mathfrak{I}\left\{\left(\langle\hat{n}_{0}\hat{b}_{0}\rangle-\langle\hat{n}_{0}\rangle\langle\hat{b}_{0}\rangle\right)\langle\hat{b}^{\dagger}_{1}\rangle\right\}\\ \nonumber
&-16|\tau r|\tau ^2\mathfrak{I}\left\{\langle\hat{b}_{0}\rangle\left(\langle\hat{b}^{\dagger}_{0}\hat{n}_{1}\rangle-\langle\hat{n}_{1}\rangle\langle\hat{b}^{\dagger}_{1}\rangle\right)\right\}.
\end{align}
}

Where the QFI corresponding to parameter $i$ is simply the diagonal element of the QFIM \cite{Liu2020}, as expressed by
\begin{equation}
\mathcal{F}_{aa} = 4 \left( \langle \partial_a \Psi | \partial_a \Psi \rangle - \left| \langle \partial_a \Psi | \Psi \rangle \right|^2 \right)
\end{equation}
 
It is evident that the single parameter QFI, denoted as $\mathcal{F}^{(a)}$, can be formulated in terms of the coefficients of the QFIM as follows.

\begin{equation} \label{Fa}
\mathcal{F}^{(a)} = \mathcal{F}_{\text{dd}} + \mathcal{F}_{\text{ss}} - 2\mathcal{F}_{\text{sd}} 
 \end{equation} 

In a context resembling the initial scenario, where the primary focus is on estimating a single parameter, the problem is modeled through a unitary operation as 
\begin{equation}
    U(\varphi) = e^{i \varphi/2} (\hat{n}_2 - \hat{n}_3).
\end{equation}
The associated QFI is characterized explicitly as

\begin{equation}
\mathcal{F}^{(b)} = \Delta^2 \hat{n}_2 + \Delta^2 \hat{n}_3
\end{equation}

Similarly, in equation (37), we obtain
{\small
\begin{align} \nonumber \label{Fb}
& \mathcal{F}(b) = \tau ^4 + |r|^4 (\Delta^2 \hat{n}_0 + \Delta^2 \hat{n}_1) + 2|\tau r|^2 \langle \hat{n}_0 \rangle + \langle \hat{n}_1 \rangle  \\ \nonumber
& + 2 \langle \hat{n}_0 \rangle \langle \hat{n}_1 \rangle - |\langle b^\dagger_0 \rangle|^2|\langle b^\dagger_1 \rangle|^2  \\ \nonumber
&-2|\tau r|^2\left(\langle\hat{b}^{2}_{0}\rangle\langle(\hat{b}^{\dagger}_{1})^{2}\rangle+\langle(\hat{b}^{\dagger}_{0})^{2}\rangle\langle\hat{b}^{2}_{1}\rangle-\langle\hat{b}_{0}\rangle^2\langle\hat{b}^{\dagger}_{1}\rangle^2-\langle\hat{b}^{\dagger}_{0}\rangle^2\langle\hat{b}_{1}\rangle^2\right)\\\nonumber
& + 2W^*r\tau ^2 - |r|^2 \left(\langle b^\dagger_0 \hat{n}_0 \rangle - \langle b^\dagger_0 \rangle\langle \hat{n}_0 \rangle\right) \langle b^\dagger_1 \rangle \\ \nonumber
& - 2W^*r\tau ^2 - |r|^2 \left(\langle \hat{n}_0 b_0 \rangle - \langle \hat{n}_0 \rangle\langle b_0 \rangle\right) \langle b^\dagger_1 \rangle \\ \nonumber
& + 2W^*r\tau ^2 - |r|^2 \langle b^\dagger_0 \rangle \left(\langle b^\dagger_{11} \hat{n}_1 \rangle - \langle b^\dagger_1 \rangle\langle \hat{n}_1 \rangle\right) \\ 
& - 2W^*r\tau ^2 - |r|^2 \langle b^\dagger_0 \rangle \left(\langle \hat{n}_1 b_1 \rangle - \langle \hat{n}_1 \rangle\langle b_1 \rangle\right).
\end{align}
 }
 
This implies that the QCRB
\begin{equation}
\Delta \varphi_{QCRB}^{(b)}= \frac{1}{\sqrt{\mathcal{F}^{(b)}}}.
\end{equation}

In the last case, we consider the most general scenario where an interferometer's upper and lower arms contain a phase shift\cite{Mishra2022}, denoted by $\varphi_1$ and $\varphi_2$, respectively. According to the literature \cite{Jarzyna2012, Lang2013, Takeoka2017}, a two-parameter estimation technique can be used to avoid the problem of counting supplementary resources, such as an external phase reference, that is unavailable. In the case when there is no external phase reference available, we are only interested in the phase shift difference $\varphi_d= \varphi_1 - \varphi_2$. Therefore, writing QFIM based on $\varphi_{s/d} = \varphi_1 \pm \varphi_2$ is more convenient. To estimate the values of $\varphi_s$ and $\varphi_d$, as seen in (\ref{QFIM}).

Where
\begin{equation} \label{Fij}
    \mathcal{F}_{ij}=4Re\{\langle\partial_{i}\Psi|\partial_{j}\Psi\rangle-\langle\partial_{i}\Psi|\Psi\rangle\langle\Psi|\partial_{j}\Psi\rangle\},
\end{equation}

The subscripts $i$ and $j$ correspond to $\varphi_s$ and $\varphi_d$.

We consider the wavevector $|\Psi\rangle$, which is expressed as
\begin{equation} 
|\Psi\rangle = e^{-i\left(\frac{\hat{n}_2 - \hat{n}_3}{2}\right) \varphi_d} e^{-i\left(\frac{\hat{n}_2 + \hat{n}_3}{2}\right) \varphi_s} |\psi_{23}\rangle
\end{equation} 
Where $\hat{n}_l = \hat{b}^\dagger_l \hat{b}_l$ stands for the number operator that goes with port $l$.By applying the previous field operator transformations \cite{Preda2019} to the input state $|\Psi\rangle$, we can obtain the state $|\psi_{23}\rangle$.

In multiparameter estimation\cite{Ataman2018}, the QCRB is provided as
\begin{equation}   \label{cov}
\text{Cov}(\hat{\varphi}) \geq \frac{1}{M}\mathcal{F}^{-1}
\end{equation} 
where $\mathcal{F}$ is the QFIM from equation (\ref{QFIM}), and $\text{Cov}(\hat{\varphi})$ is the estimator's covariance matrix, which includes both $\varphi_d$ and $\varphi_s$. Whose elements are
\begin{equation} 
\text{Cov}(\hat{\varphi})_{ij} = E(\hat{\varphi}_i \hat{\varphi}_j) - E(\hat{\varphi}_i)E(\hat{\varphi}_j),
\end{equation} 

Here, $E$ is a mathematical expectation. We consider $N = 1$ for the balance of the paper. Specifically, when taking phase difference sensitivity into account, we've got
\begin{equation} 
(\Delta \varphi_d)^2 \geq (\mathcal{F}^{-1})_{dd}
\end{equation} 
Where the expression for the first diagonal element of the matrix $\mathcal{F}^{-1}$ inverse is as follows
\begin{equation} \label{QFI 2p}
\mathcal{F}(c) = \frac{1}{{(\mathcal{F}^{-1})_{dd}}} = \mathcal{F}_{dd} - \frac{(\mathcal{F}_{sd})^2}{\mathcal{F}_{ss}}
\end{equation} 

Inequality  (\ref{cov}) can therefore be saturated, implying the two-parameter QCRB.
\begin{equation} \label{QCRB(c)}
\Delta \varphi_{QCRB}^{(c)}= \frac{1}{\sqrt{\mathcal{F}^{(c)}}}.
\end{equation}
Using the definition in equation  (\ref{Fij}), the elements of QFIM (\ref{QFIM}), namely $ \mathcal{F}_{ss}$, $ \mathcal{F}_{dd}$, and $\mathcal{F}_{sd}$, can be determined as follows
\begin{widetext}
\begin{align} 
\mathcal{F}_{ss}=&\Delta^2\hat{n}_{0}+\Delta^2\hat{n}_{1},\\
\mathcal{F}_{dd}=&\left(\tau ^2-|r|^2\right)^2\left(\Delta^2\hat{n}_{0}+\Delta^2\hat{n}_{1}\right)+8|\tau r|^2\left(
\langle\hat{n}_{0}\rangle\langle\hat{n}_{1}\rangle-|\langle\hat{b}_{0}\rangle|^2|\langle\hat{b}_{1}\rangle|^2-\mathfrak{r}\left\{\langle(\hat{b}_{0}^{\dagger})^2\rangle\langle\hat{b}_{1}^2\rangle-\langle\hat{b}_{0}^{\dagger}\rangle^2\langle\hat{b}_{1}\rangle^2\right\}\right)\notag\\
&+4|\tau r|^2\left(\langle\hat{n}_{0}\rangle+\langle\hat{n}_{1}\rangle\right)-8|\tau r|\left(\tau ^2-|r|^2\right)\left(\mathfrak{I}\left\{\left(\langle\hat{b}_{0}^{\dagger}\hat{n}_{0}\rangle-\langle\hat{b}_{0}^{\dagger}\rangle\langle\hat{n}_{0}\rangle\right)\langle\hat{b}_{1}\rangle+\langle\hat{b}_{0}\rangle\left(\langle\hat{b}_{1}^{\dagger}\hat{n}_{1}\rangle-\langle\hat{n}_{1}\rangle\langle\hat{b}_{1}^{\dagger}\rangle\right)\right\}\right),\\
\mathcal{F}_{sd}=&\left(\tau ^2-|r|^2\right)\left(\Delta^2\hat{n}_{0}-\Delta^2\hat{n}_{1}\right)+4|\tau r|\mathfrak{I}\left\{\langle\hat{b}_{0}\rangle\langle\hat{b}^{\dagger}_{1}\rangle-\left(\langle\hat{n}_{0}\hat{b}_{0}\rangle-\langle\hat{n}_{0}\rangle\langle\hat{b}_{0}\rangle\right)\langle\hat{b}^{\dagger}_{1}\rangle+\langle\hat{b}_{0}\rangle\left(\langle\hat{b}_{1}^{\dagger}\hat{n}_{1}\rangle-\langle\hat{b}^{\dagger}_{1}\rangle\langle\hat{n}_{1}\rangle\right)\right\},
\end{align}
\end{widetext}

In which the variance is expressed as $\Delta^2n$, which can be defined as $\Delta^2\hat{n} = \langle\hat{n}^2\rangle - \langle\hat{n}\rangle^2$
 If $\mathcal{F}_{sd} = \mathcal{F}_{ss}$, then the above equation and the two-parameter QFI (\ref{QFI 2p}) are equal; $\mathcal{F}^{(a)} = \mathcal{F}^{(c)}$ We can prove that.
 
 \begin{equation} 
  \mathcal{F}^{(a)}\geq \mathcal{F}^{(c)}
\end{equation} 

\section{Unveiling Entangled SU(2) Coherent States and Quantum Fisher Information in Mach-Zehnder Interferometers}

This section focuses on input states characterized by entangled SU(2) coherent states combined with a vacuum state. Specifically, we consider the pure two-mode input state\cite{Berrada}; this state is conceptualized as a superposition of macroscopic spin coherent states, denoted as
\begin{equation}
|\Psi_{\text{in}}\rangle = |\xi_i, k\rangle_i
\end{equation}

Then the expressions of $ \mathcal{F} _{ss}$, $ \mathcal{F} _{dd}$, and $\mathcal{F} _{sd}$ are redefined in terms of the second moment of the photon number operator $\Delta^2 n_1$ as follows

\begin{equation}
\begin{aligned}
& \mathcal{F} _{s s}=\Delta^2 n_1=\left\langle n_1^2\right\rangle-\left\langle n_1\right\rangle^2 \\
& \mathcal{F} _{d d}=\left(\tau ^2-|r|^2\right)^2 \Delta^2 n_1+4|\tau  r|^2\left\langle n_1\right\rangle \\
&  \mathcal{F} _{s d}=-\left(\tau ^2-|r|^2\right) \Delta^2 n_1
\end{aligned}
\end{equation}

These expressions provide a quantitative measure of the fluctuations in the photon number. 

\begin{figure}[ht] 
	\centering
	\includegraphics[scale=0.55]{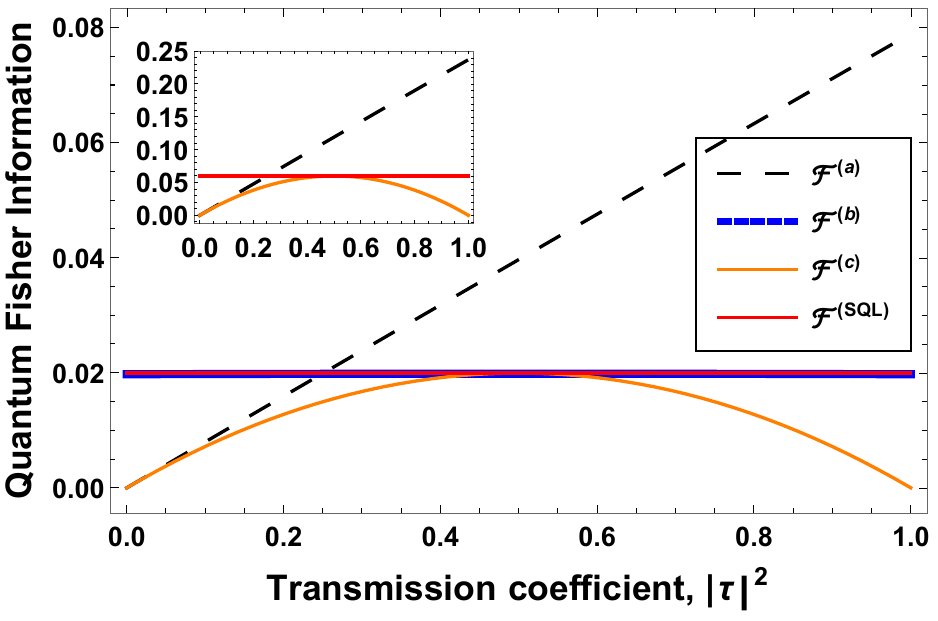}
	\caption{Quantifying Quantum Information: Comparing QFI Dynamics with SQL Across Diverse Scenarios with $j=1$ and $j=3$ Illustrated in a Small Inset Graph at the Top Left Corner.} \label{QFI in Diverse Scenarios}
\end{figure}
  
We also represent quantum Fisher information (QFI) as $F_i$. By employing the outcomes of the three QFIs mentioned in the preceding section—namely, $\mathcal{F}^{(a)}$, $\mathcal{F}^{(b)}$, and $\mathcal{F}^{(c)}$, derived from equations (\ref{Fa}), (\ref{Fb}), and (\ref{QFI 2p})—we can readily confirm that the QFI in our initial states for the last three scenarios referred to in the second section has been obtained.In such a way that the expressions for $\mathcal{F} ^{(i)}$ and $\mathcal{F} ^{(ii)}$ offer insights into the QFI under different conditions.
\begin{equation}
\begin{aligned} 
&\mathcal{F} ^{(i)}=4\tau ^4 \Delta_{n_1}^2+\mathcal{F} ^{(2 p)} \\
& F^{(ii)}=\left(\tau ^4+|r|^4\right) \Delta_{n_1}^2+2|\tau  r|^2\left\langle n_1\right\rangle 
\end{aligned}
\end{equation}

Moreover, $\mathcal{F} ^{(c)}$ is expressed as $\mathcal{F} _{d d}-\frac{\left(\mathcal{F} _{sd}\right)^2}{\mathcal{F} _{ss}}$, revealing its dependence on the components $\mathcal{F} _{dd}$,$F_{sd}$ and $\mathcal{F} _{ss}$, then we obtained 
\begin{equation}
\mathcal{F} ^{(2 p)}=4|\tau  r|^2\left\langle n_1\right\rangle 
\end{equation}

Fig.\ref{QFI in Diverse Scenarios} illustrates the dynamic behavior of three QFI metrics, denoted as $\mathcal{F}^{(a)}$, $\mathcal{F}^{(b)}$, and $\mathcal{F}^{(c)}$, concerning the transmission coefficient $\tau^2$ of the initial BS, along with an additional metric denoted as $\mathcal{F}^{(\text{SQL})}$ representing the standard quantum limit (SQL). This analysis focuses on coherent states as the single input state in both cases, firstly for $j=1$, and secondly, in the smaller inset graph in the top left corner, representing  $j=3$. The metric $\mathcal{F}^{(a)}$, represented by the black line, shows a linear increase from 0 at $\tau=0$ to $2k\sinh^2(v)$ at $\tau=1$, thereby surpassing the SQL, indicating a significant improvement over the standard measurement. In contrast, the metric $\mathcal{F}^{(b)}$, represented by the blue graph, remains relatively stable across different values of the transmission coefficient $\tau$, reaching the SQL. The metric $\mathcal{F}^{(c)}$, represented by the orange graph, peaks when the system is balanced, specifically when $\tau=|r|=1/\sqrt{2}$, and drops to its minimum value of 0 at the extremes, also reaching the standard quantum limit under optimal conditions.\par

In summary, Fig.\ref{QFI in Diverse Scenarios} shows that for scenarios (b) and (c), the QFI values reach the SQL, whereas for scenario (a), the QFI surpasses this limit, indicating a significant improvement over the standard measurement. Ultimately, this section comprehensively analyzes Quantum Fisher Information and its dependencies on different parameters. It offers valuable insights into the precision of phase estimation in the context of the considered SU(2) interferometer and input states\cite{Berrada}.

\section {Discovering Realistic Detection Schemes while Exploring Phase Sensitivity in Mach-Zehnder Interferometers}\label{sec5}

The next step will be using a beam splitter (BS2) to close the MZI. $ \tau ' (r')$ is the transmission (reflection) coefficient of BS2. We then examine the performance of three realistic detection schemes, namely the single mode intensity(SMI), the difference intensity(DI), and the balanced homodyne (BH) detection\cite{Stefan}.
Moving on to the quantum parameter estimation problem, we introduce an experimentally accessible Hermitian operator denoted as $\hat{\mathcal{D}}$, dependent on a parameter $\varphi$. In our specific case, $\varphi$ denotes the phase shift within an MZI, a quantity that could be observable or not \cite{Ataman2018}.
Expressing the average of this operator involves
\begin{equation}
\langle\hat{\mathcal{D}}(\varphi)\rangle = \langle\Psi|\hat{\mathcal{D}}(\varphi)|\Psi\rangle
\end{equation}
Here, $\Psi$ signifies the wave function of the system. Applying a small variation $\delta\varphi$ to the parameter $\varphi$ induces a change described by
\begin{equation}
\langle\hat{\mathcal{D}}(\varphi + \delta\varphi)\rangle \approx \langle\hat{\mathcal{D}}(\varphi)\rangle + \frac{\partial\langle\hat{\mathcal{D}}(\varphi)\rangle}{\partial\varphi} \delta\varphi
\end{equation}
The experimental detectability of the difference between $\langle\hat{\mathcal{D}}(\varphi+ \delta\varphi)$ and $\langle\hat{\mathcal{D}}(\varphi)$ relies on satisfying the condition
\begin{equation}
\langle\hat{\mathcal{D}}(\varphi + \delta\varphi)\rangle - \langle\hat{\mathcal{D}}(\varphi)\rangle \geq \Delta\hat{\mathcal{D}}(\varphi)
\end{equation}
Here, $\Delta\hat{\mathcal{D}} $ is the standard deviation of $\hat{\mathcal{D}} $, defined as the square root of the variance $\Delta^2 \hat{\mathcal{D}}$, expressed as
\begin{equation}
\Delta\hat{\mathcal{D}} =\sqrt{\langle\hat{\mathcal{D}}^2\rangle - \langle\hat{\mathcal{D}}\rangle^2}
\end{equation}
If the value of $ \delta\varphi$ saturates the inequality  \cite{dAriano1994}, then this variation is termed sensitivity, denoted by
\begin{equation}
\Delta \varphi=\frac{\Delta \hat{\mathcal{D}}}{\left|\frac{\partial}{\partial \varphi}\langle\hat{\mathcal{D}}\rangle\right|} 
\end{equation}
In the subsequent discussion, $ \varphi$ represents the total phase shift inside the interferometer, divided into two parts: $ \varphi_i $, the quantity to be measured, and $\varphi_{exp}$, experimentally controllable. The relationship is expressed as
\begin{equation}
\varphi = \varphi_i + \varphi_{\text{exp}}.
\end{equation}

In interferometry, the magnitude of the unknown phase shift $\varphi_i$ must be significantly smaller than that of the total phase shift $\varphi$. This ensures that $\varphi_i$ has minimal impact on $\varphi$. Consequently, the experimenter must fine-tune $\varphi_{exp}$ to approximate the optimal phase shift $\varphi_{opt}$ for optimal performance\cite{Mishra2022}.

Now, considering the various detection schemes employed in interferometric measurements, such as single-mode intensity detection, difference intensity detection, and balanced homodyne detection, each scheme's efficacy is intimately tied to its sensitivity to phase variations. These sensitivities encapsulate how effectively each detection method can discern phase changes, thereby shedding light on the underlying physical phenomena being studied. Determining these sensitivities offers valuable insights into each detection scheme's performance characteristics and limitations, guiding experiment design and data interpretation.
These sensitivities can be calculated as follows
\subsection{Single-mode intensity detection scheme}
In the SMI detection scheme, we focus on a sole photocurrent located at output port 4. This is represented by its corresponding operator, denoted as $\hat{n}_{4}=\hat{b}_{4}^{\dagger}\hat{b}_{4}$.
 The phase sensitivity in this scenario is defined as
\begin{equation}
    \Delta\varphi_{sing}=\frac{\Delta\hat{n}_{4}}{|\frac{\partial}{\partial\varphi}\langle\hat{n}_{4}\rangle|}.
\end{equation}
From equation (\ref{transformations22}), it is possible to find the average number of photons relative to the input field operator
{\small
\begin{align}
    \langle\hat{n}_{4}\rangle=&\left(|\tau  \tau'|^{2}+|rr'|^{2}-2|\tau \tau ' rr'|\cos\varphi\right)\langle\hat{n}_{0}\rangle\\ \nonumber
    &+\left(|\tau  r'|^{2}+|\tau ' r|^{2}+2|\tau \tau ' rr'|\cos\varphi\right)\langle\hat{n}_{1}\rangle\\ \nonumber
    &+\left( \tau ^{\ast}r( | \tau '|^{2}-|r'|^{2})+ \tau '^{\ast}r'(|\tau | ^{2}e^{-i\varphi}-|r|^{2}e^{i\varphi})\right)\langle\hat{b}_{0}^{\dagger}\rangle\langle\hat{b}_{1}\rangle\\ \nonumber
    &+\left( \tau ^{\ast}r(|r'|^{2}- | \tau '|^{2})+ \tau '^{\ast}r'(|r|^{2}e^{-i\varphi}-|\tau | ^{2}e^{i\varphi})\right)\langle\hat{b}_{0}\rangle\langle\hat{b}_{1}^{\dagger}\rangle.
\end{align}
}
Using the above equation, we immediately get
\begin{align}
\frac{\partial\langle\hat{n}_{4}\rangle}{\partial\varphi}=&2|\tau \tau ' rr'|\sin\varphi(\langle\hat{n}_{0}\rangle-\langle\hat{n}_{1}\rangle)\\
&+2| \tau '^{\ast}r'|\mathfrak{r}\left\{\left(|\tau | ^{2}e^{-i\varphi}+|r|^{2}e^{i\varphi}\right)\langle\hat{b}_{0}^{\dagger}\rangle\langle\hat{b}_{1}\rangle\right\}.\nonumber
\end{align}

To obtain $\Delta^{2}\hat{n}_{4}$, we initially compute the operator $\hat{n}_{4}$ squared. Subsequently, we derive the ultimate expression for $\Delta^{2}\hat{n}_{4}$ as follows

\small{
\begin{align}\nonumber
\Delta^{2}\hat{n}_{4}=&A_{0}^{2}\Delta^{2}\hat{n}_{0}+A_{1}^{2}\Delta^{2}\hat{n}_{1}+2|A_{01}^{2}|\mathfrak{r}\left\{\langle\hat{b}_{0}^{2}\rangle\langle(\hat{b}_{1}^{\dagger})^{2}\rangle-\langle\hat{b}_{0}\rangle^{2}\langle\hat{b}_{1}^{\dagger}\rangle^{2}\right\}\\ \nonumber 
&+|A_{01}|^{2}\left(\langle\hat{n}_{0}\rangle+\langle\hat{n}_{1}\rangle+2\langle\hat{n}_{0}\rangle\langle\hat{n}_{1}\rangle-2|\langle\hat{b}_{0}\rangle|^{2}|\langle\hat{b}_{1}\rangle|^{2}\right)\\ \nonumber
&+2A_{0}\mathfrak{r}\left\{A_{01}(\langle\hat{n}_{0}\hat{b}_{0}^{\dagger}\rangle+\langle\hat{b}_{0}^{\dagger}\hat{n}_{0}\rangle-2\langle\hat{n}_{0}\rangle\langle\hat{b}_{0}^{\dagger}\rangle)\langle\hat{b}_{1}\rangle\right\}\\ \nonumber
&+2A_{1}\mathfrak{r}\left\{A_{01}\langle\hat{b}_{0}^{\dagger}\rangle(\langle\hat{n}_{1}\hat{b}_{1}\rangle+\langle\hat{b}_{1}\hat{n}_{1}\rangle-2\langle\hat{n}_{1}\rangle\langle\hat{b}_{1}\rangle)\right\},
\end{align}
}

Where
\begin{align} \label{A1}
A_{0}=&|\tau  \tau'|^{2}+|rr'|^{2}-2|\tau \tau ' rr'|\cos\varphi,\\
A_{1}=&|\tau  r'|^{2}+|\tau ' r|^{2}+2|\tau \tau ' rr'|\cos\varphi,\\
A_{01}=& \tau ^{\ast}r(2 | \tau '|^{2}-1)+ \tau '^{\ast}r'(|\tau | ^{2}e^{-i\varphi}-|r|^{2}e^{i\varphi}).
\end{align}

\subsection{Difference-intensity detection scheme}

In the DI detection scheme, which is only sensitive to the difference between the phase shifts $\varphi_{1}$ and $\varphi_{2}$, we calculate the difference between the output photocurrents, denoted as $N_{d}$, specifically those detected at D4 and D5, as illustrated in FIG \ref{detection schemes}. Hence, the output operator $N_{d}$ is delineated as
\begin{equation}
    N_{d}=\hat{b}^{\dagger}_{4}\hat{b}_{4}-\hat{b}^{\dagger}_{5}\hat{b}_{5}.
\end{equation}

To express the operator $N_{d}$ in terms of the input field operators, we need the field operator transformations,

\begin{equation} \label{transformations2}
    \hat{b}_{4}=  \tau '\hat{b}_{2}+r'\hat{b}_{3},\quad \hat{b}_{5}=r'\hat{b}_{2}+  \tau '\hat{b}_{3},
\end{equation}
In this context, $ \tau'(r')$ denotes the transmission (reflection) coefficients of the second beamsplitter (BS2). Through the application of the field operator transformations detailed in  (\ref{transformations1}), we derive the following result

\begin{align} \label{transformations22}
    &\hat{b}_{4}=e^{-i\varphi_{2}}[(\tau \tau '+rr'e^{-i\varphi})\hat{b}_{0}+(  \tau  r 'e^{-i\varphi}+  r \tau  ')\hat{b}_{1}],\\
    &\hat{b}_{5}=e^{-i\varphi_{2}}[(  \tau  r '+  r \tau  'e^{-i\varphi})\hat{b}_{0}+(  \tau  \tau 'e^{-i\varphi}+rr')\hat{b}_{1}],
\end{align}

Here, $\varphi=\varphi_{1}-\varphi_{2}$ is defined as the difference between $\varphi_{1}$ and $\varphi_{2}$.\\\\
The expression for $N_{d}$ can be finalized. By substituting the field operator transformations into the definition of $N_{d}$, we get
{\small
\begin{align}
    N_{d}=&\left[(|\tau | ^{2}-|r|^{2})( | \tau '|^{2}-|r'|^{2})-4|\tau \tau ' rr'|\cos\varphi\right](\hat{n}_{0}-\hat{n}_{1})\\ \nonumber
    &+2\left( \tau ^{\ast}r(|r'|^{2}- | \tau '|^{2})+(|r|^{2}e^{-i\varphi}-|\tau | ^{2}e^{i\varphi}) \tau '^{\ast}r'\right)\hat{n}_{0}\hat{n}_{1}^{\dagger}\\ \nonumber
    &+2\left( \tau ^{\ast}r( | \tau '|^{2}-|r'|^{2})+(|\tau | ^{2}e^{-i\varphi}-|r|^{2}e^{i\varphi}) \tau '^{\ast}r'\right)\hat{n}^{\dagger}_{0}\hat{n}_{1}.
\end{align}
}
The phase sensitivity is defined as
\begin{equation}
    \Delta\varphi_{df}=\frac{\Delta\hat{N}_{d}}{|\frac{\partial}{\partial\varphi}\langle\hat{N}_{d}\rangle|}.
\end{equation}
Where the derivative of $\hat{N}_{d}$ with respect to $\varphi$ is given by
\begin{align}
    \frac{\partial}{\partial\varphi}\langle\hat{N}_{d}\rangle=&4|\tau  \tau 'rr'|\sin\varphi(\langle\hat{n}_{0}\rangle-\langle\hat{n}_{1}\rangle)\\
    &+4|  \tau 'r'|\mathfrak{r}\{(|r|^{2}e^{-i\varphi}+\tau ^{2}e^{i\varphi})\langle\hat{b}_{0}\rangle\langle\hat{n}^{\dagger}_{1}\rangle\}
\end{align}
Therefore, the variance of the operator $\hat{N}_{d}$ can be expressed by the following form

{\footnotesize{
\begin{align}
    \Delta^{2}\hat{N}_{d}&=A_{d}^{2}(\Delta^{2}\hat{n}_{0}+\Delta^{2}\hat{n}_{1})+|C_{d}|^{2}(\langle\hat{n}_{0}\rangle+\langle\hat{n}_{1}\rangle)\\ \nonumber
    &+2|C_{d}|^{2}(\langle\hat{n}_{0}\rangle\langle\hat{n}_{1}\rangle-|\langle\hat{b}_{0}\rangle|^{2}||\langle\hat{b}_{1}\rangle|^{2})\\ \nonumber
    &+2\mathfrak{r}\{C_{d}^{2}(\langle\hat{b}_{0}^{2}\rangle\langle(\hat{b}_{1}^{\dagger})^{2}\rangle)-\langle\hat{b}_{0}\rangle^{2}\langle\hat{b}_{1}^{\dagger}\rangle^{2}\}\\ \nonumber
    &+4A_{d}\mathfrak{r}\left\{C_{d}\left((\langle\hat{n}_{0}\hat{b}_{0}\rangle-\langle\hat{n}_{0}\rangle\langle\hat{b}_{0}\rangle)\langle\hat{b}_{1}^{\dagger}\rangle-\langle\hat{b}_{0}\rangle(\langle\hat{b}_{1}^{\dagger}\hat{n}_{1}\rangle-\langle\hat{n}_{1}\rangle\langle\hat{b}_{1}^{\dagger}\rangle)\right)\right\} 
\end{align}
}}

Where
\begin{align} \label{Ad and Cd}
A_{d}=&1-2\left(\tau |r'|+|r||  \tau '|\right)^{2}+4|\tau r||  \tau 'r'|(1-\cos\varphi),\\
C_{d}=&2|  \tau 'r'|\sin\varphi+2i\left(|\tau r|(1-2 | \tau '|^{2})+(1-2\tau ^{2})|  \tau 'r'|\cos\varphi\right).
\end{align}
Which satisfy the following conditions
\begin{equation}
    A_{d}^{2}+|C_{d}|^{2}=1.
\end{equation}
In this detection scheme, the phase sensitivity remains the same for both scenarios (i) and (ii). To simplify, we will collectively denote the phase sensitivity in all detection schemes as $\Delta\varphi_{df}$.

\subsection{Balanced homodyne detection scheme}

Now, we consider the BH detection scheme, a detector with access to an external phase reference \cite{Genoni2011}; in a typical Mach-Zehnder interferometer, an external phase reference can be introduced by mixing the input beams with a reference beam. The relative phase shift between the arms of the interferometer can then be accurately measured, enhancing the precision of phase estimation beyond the standard quantum limit. Similarly,  Including an external reference beam aligns the phase shifts with this reference, allowing pure states without phase averaging\cite{Jarzyna 2012, Stefan}. This approach yields higher QFI values, implying better phase estimation precision \cite{ Zhang 2020, Adhikari 2019}, specifically in this paper focusing on output port 4, as illustrated in FIG \ref{detection schemes}. Thus, the operator of interest for modeling this detection scheme is expressed as
\begin{equation}
    \hat{Y}_{\varphi_{L}}=\mathfrak{r}\left\{e^{-i\varphi_{L}}\hat{b}_{4}\right\}
\end{equation}
Here, $\varphi_L$ denotes the phase of the local coherent source $|\gamma\rangle$, where $|\gamma\rangle = |\gamma|e^{i\varphi_L}$, and $\gamma$ is a complex number. In this detection scheme, the phase sensitivity is defined as follows

\begin{equation}
    \Delta\varphi_{hom}=\frac{\sqrt{\Delta^{2}\hat{Y}_{\varphi_{L}}}}{|\frac{\partial\langle\hat{Y}_{\varphi_{L}}\rangle}{\partial\varphi}|}.
\end{equation}
With the utilization of the field operator transformations (\ref{transformations22}), we can find the final expression of $\langle\hat{Y}_{\varphi_{L}}\rangle$
\begin{align}
    \langle\hat{Y}_{\varphi_{L}}\rangle=&\mathfrak{r}\left\{e^{-i\varphi_{L}}\left((  \tau  \tau 'e^{-i\varphi_{2}}+rr'e^{-i\varphi_{1}})\langle\hat{b}_{0}\rangle \right.\right.\\
    &\left.\left. +(  \tau  r 'e^{-i\varphi_{1}}+  r \tau  'e^{-i\varphi_{2}})\langle\hat{b}_{1}\rangle\right)\right\}, \nonumber
\end{align}
and the variance of the above operator is given by
\begin{align}
\Delta^{2}\hat{Y}_{\varphi_{L}}=&\frac{1}{4}+2\mathfrak{r}\left\{K_{0}^{2}\Delta^{2}\hat{b}_{0}+K_{1}^{2}\Delta^{2}\hat{b}_{1}\right\}\\
&+2|K_{0}|^{2}(\langle\hat{n}_{0}\rangle-|\langle\hat{b}_{0}\rangle|^{2})+2|K_{1}|^{2}(\langle\hat{n}_{1}\rangle-|\langle\hat{b}_{1}\rangle|^{2}), \nonumber
\end{align}
where the coefficients $K_{0}$ and $K_{1}$ are defined as
\begin{align} \label{A and B}
K_{0}=&\frac{1}{2}e^{-i(\varphi_{L}+\varphi_{2})}\left(  \tau  \tau '+rr'e^{-i\varphi}\right),\\
K_{1}=&\frac{1}{2}e^{-i(\varphi_{L}+\varphi_{2})}\left(  \tau  r 'e^{-i\varphi}+  \tau 'r\right).&
\end{align}
For scenario (i) in Figure 1, where $\varphi_{1}=\varphi$ and $\varphi_{2}=0$, the absolute value of $\langle\hat{Y}_{\varphi_{L}}\rangle$ with respect to $\varphi$ is determined by
\begin{equation} \label{di}
    |\frac{\partial\langle\hat{Y}_{\varphi_{L}}\rangle}{\partial\varphi}|=|\mathfrak{r}\left\{e^{-i(\varphi_{L}+\varphi)}\left(r\langle\hat{b}_{0}\rangle+  \tau \langle\hat{b}_{1}\rangle\right)\right\}||r'|
\end{equation}
and for scenario (ii), where $\varphi_{1}=-\varphi_{2}=\varphi/2$, we obtain
\begin{align} \label{dii}
    |\frac{\partial\langle\hat{Y}_{\varphi_{L}}\rangle}{\partial\varphi}|=&\frac{1}{2}|\mathfrak{r}\left\{ie^{-i\varphi_{L}}\left(\left(  \tau  \tau 'e^{i\varphi/2}-rr'e^{-i\varphi/2}\right)\langle\hat{b}_{0}\rangle\right.\right.\\
     &\left.\left. +\left(  r \tau  'e^{i\varphi/2}-  \tau  r 'e^{-i\varphi/2}\right)\right)\langle\hat{b}_{1}\rangle\right\}|
\end{align}
\\

\section {Comparative Analysis of Phase Sensitivities for Different Detection Schemes in SU(2) Coherent States}

In this section, we analyze the phase sensitivities achievable by the three considered detection schemes, as detailed in section \ref{sec5}, for input $SU(2)$ coherent states with the QCRBs implied by the various quantum Fisher information (QFIs) discussed in section \ref{section 3}.

Utilizing the results of the phase sensitivities reported in the preceding section, specifically:
$\Delta\varphi_{sing}$,
$\Delta\varphi_{d}$, and $\Delta\varphi_{hom}$, and taking into account the input state (\ref{input state}), it is easy to verify that the phase sensitivities in our input states  can be expressed as follows\\\\

For an SMI detection scheme\cite{Mishra2022}, we obtain the phase sensitivity in all considered scenarios as
\begin{equation}
\Delta\varphi_{sing}=\frac{\Delta_{i}\hat{n}_{4}}{2|\tau  \tau 'rr'||\sin\varphi \langle\hat{n}_{1}\rangle|}
\end{equation}
Alternatively, the analytical expression for the phase sensitivity of our input SU(2) coherent states is provided in the following form
{\small
\begin{align}
    \Delta\varphi_{sing}=&\frac{\sqrt{A_{1}^{2}\Delta^{2}\hat{n}_{1}+|A_{01}|^{2}\langle\hat{n}_{1}\rangle}}{2|\tau  \tau 'rr'||\sin\varphi \langle\hat{n}_{1}\rangle|}
\end{align}
}
where $A_{1}$ and $A_{01}$ are explicitly provided in equation  (\ref{A1}).\\\\

For a DI detection scheme, we obtain the final analytical expression for the phase sensitivity of SU(2) coherent states as
\begin{equation}
    \Delta\varphi_{d}=\frac{\Delta_{i}\hat{N}_{d}}{4| \tau  r   \tau 'r'||\sin\varphi \langle\hat{n}_{1}\rangle|},
\end{equation}
From the above expression of the phase sensitivity, we calculate its final analytical expression for the SU(2) coherent state type.
\begin{align}
    \Delta\varphi_{d}=&\frac{\sqrt{A_{d}^{2}\Delta^{2}\hat{n}_{1}+|C_{d}|^{2}\langle\hat{n}_{1}\rangle}}{4| \tau  r     \tau 'r'||\sin\varphi \langle\hat{n}_{1}\rangle|}    
\end{align}

In both scenarios (i) and (ii), we achieve the same result for the phase sensitivity in this detection scheme, where $A_{d}$ and $C_{d}$ are defined in equation (\ref{Ad and Cd}).\\\\
For a BH detection scheme, we have
\begin{equation}
\Delta^{2}\hat{Y}_{\varphi_{L}}=\frac{1}{4}+2\mathfrak{r}\left\{K_{1}^{2}\Delta^{2}\hat{b}_{1}\right\}+2|K_{1}|^{2}(\langle\hat{n}_{1}\rangle-|\langle\hat{b}_{1}\rangle|^{2})
\end{equation}
In the case of scenario (a), where $\varphi_{1}=\varphi$ and $\varphi_{2}=0$, and assuming $\varphi_{L}=\varphi$, the variance of the operator $\hat{Y}_{\varphi_{L}}$ is given by
{\small
\begin{align}
    \Delta^{2}\hat{Y}_{\varphi_{L}}=&\frac{1}{4}-\frac{1}{2\tanh^{2}{\frac{\theta}{2}}}\left(|\tau  r'|^{2}\cos2\varphi+|\tau ' r|^{2}+2|\tau  \tau 'rr'|\cos\varphi\right)\mu \\
    &+\frac{1}{2}\left(|\tau  r'|^{2}+|\tau ' r|^{2}+2|\tau  \tau 'rr'|\cos\varphi\right)\left(	\bar{n}-|\langle\nu\rangle|^{2}\right)
\end{align}
}
Where
 {\small
\begin{align} 
\mu=&|C(|\lambda|)|^{2}\sum_{n=2}^{2r}\frac{|\lambda|^{2n}}{(n-2)!(2r-n)!\sqrt{(2r-n+2)(2r-n+1)}}\\ \nonumber
&-|C(|\lambda|)|^{4}\left(\sum_{n=1}^{2r}\frac{|\lambda|^{2n}}{(n-1)!(2r-n)!\sqrt{2r-n+1}}\right)^{2}\\
\nu=&\frac{|C(|\lambda|)|^{2}}{\tanh(\theta/2)}\sum_{n=1}^{2r}\frac{|\lambda|^{2n}}{(n-1)!(2r-n)!\sqrt{2r-n+1}}\\
\bar{n}=&C(|\lambda|)^{2} \sum_{n=0}^{2r} \frac{n|\lambda|^{2n}}{n!(2r-1)!}.
\end{align}
}
From equation (\ref{di}), we obtain

\begin{equation}
    |\frac{\partial\langle\hat{Y}_{\varphi_{L}}\rangle}{\partial\varphi}|=|\mathfrak{r}\left\{e^{-i(\varphi_{L}+\varphi)}\langle\hat{b}_{1}\rangle\right\}||\tau  r'|,
\end{equation}
Based on these findings, we can determine the level of sensitivity to phase variations.
\begin{align}
    \Delta\varphi_{hom}^{(b)}=&\frac{\sqrt{ \Delta_{i}^{2}\hat{Y}_{\varphi_{L}}}}{|\tau  r'||\cos(\varphi) \nu|}
\end{align}
In the scenario (b), where $\varphi_{1}=-\varphi_{2}=\varphi/2$, the above variance takes the following expression 
\begin{align}
    \Delta^{2}\hat{Y}_{\varphi_{L}}=&\frac{1}{4}-\frac{1}{2}\left((|\tau  r'|^{2}+|\tau ' r|^{2})\cos\varphi+2|\tau  \tau 'rr'|\right)\mu\\
    &+\frac{1}{2}\left(|\tau  r'|^{2}+|\tau ' r|^{2}+2|\tau  \tau 'rr'|\cos\varphi\right)\left(	\bar{n}-|\nu|^{2}\right)
\end{align}

From equation (\ref{dii}) we get
\begin{equation}
    |\frac{\partial\langle\hat{Y}_{\varphi_{L}}\rangle}{\partial\varphi}|=\frac{1}{2}|\mathfrak{r}\left\{ie^{-i\varphi_{L}}\left(  r \tau  'e^{i\varphi/2}-  \tau  r 'e^{-i\varphi/2}\right)\langle\hat{b}_{1}\rangle\right\}|
\end{equation}
With these findings, similar to the previous scenario, we can determine the phase sensitivity.
\begin{align}
    \Delta\varphi_{hom}^{(c)}=&\frac{2\sqrt{ \Delta_{i}^{2}\hat{Y}_{\varphi_{L}}}}{||\tau ' r|-|\tau  r'|||\cos(\frac{\varphi}{2}) \nu|}
\end{align}

The results reported in Fig.\ref{The variation of phase sensitivity and BCRB} describe the variation of the phase sensitivities as a function of the phase shift of the MZI for coherent input state. These phase sensitivities are examined in detection schemes discussed previously alongside the corresponding  QCRB(c) indicated in the equation \ref{QCRB(c)}.

We plot in Fig.\ref{The variation of phase sensitivity and BCRB}(a) the SMI and DI detection schemes where $\tau =0.9$ and $\tau'=0.1$ for BS1 and BS2, respectively, where $j=1$. In this case, it's observed that the sensitivity variation doesn't reach the Cramér-Rao limit. Additionally, the shot-noise limit SQL is included for comparison, highlighting that the sensitivity variations of the SMI and DI schemes do not achieve the SQL either.

\begin{widetext}

	\begin{figure}[ht] 
	\begin{minipage}[b]{.50\linewidth}
		\centering
		
		\includegraphics[scale=0.425]{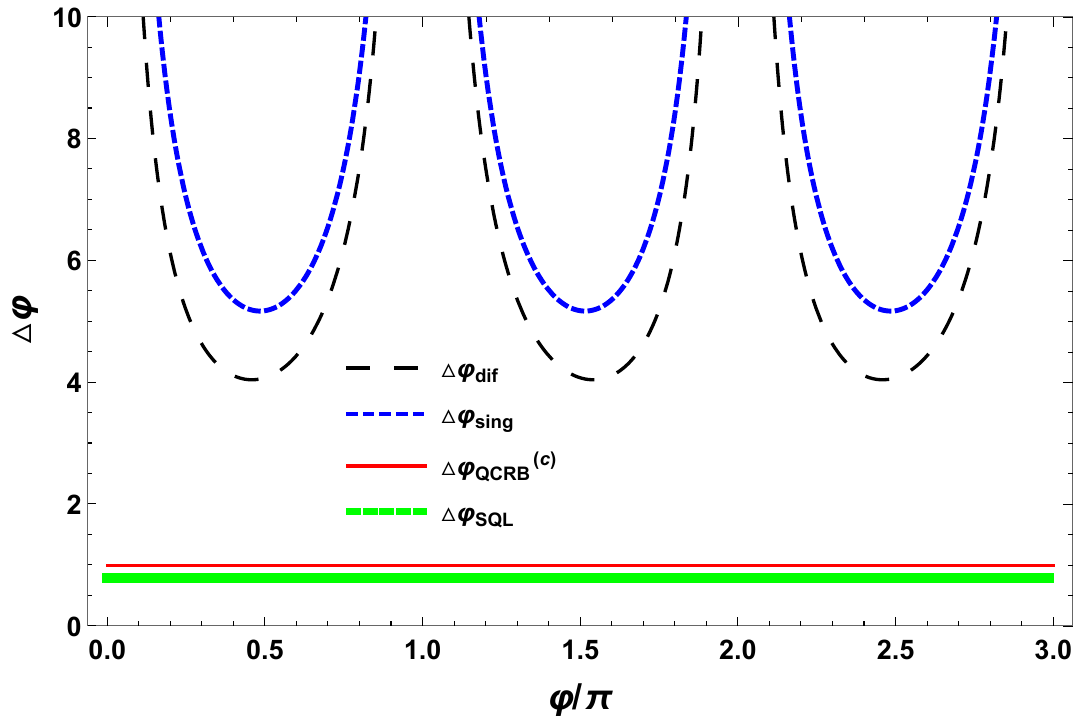} 
  \vfill $\left(a\right)$
	\end{minipage} \quad
	\begin{minipage}[b]{.45\linewidth}
		\centering
		\includegraphics[scale=0.425]{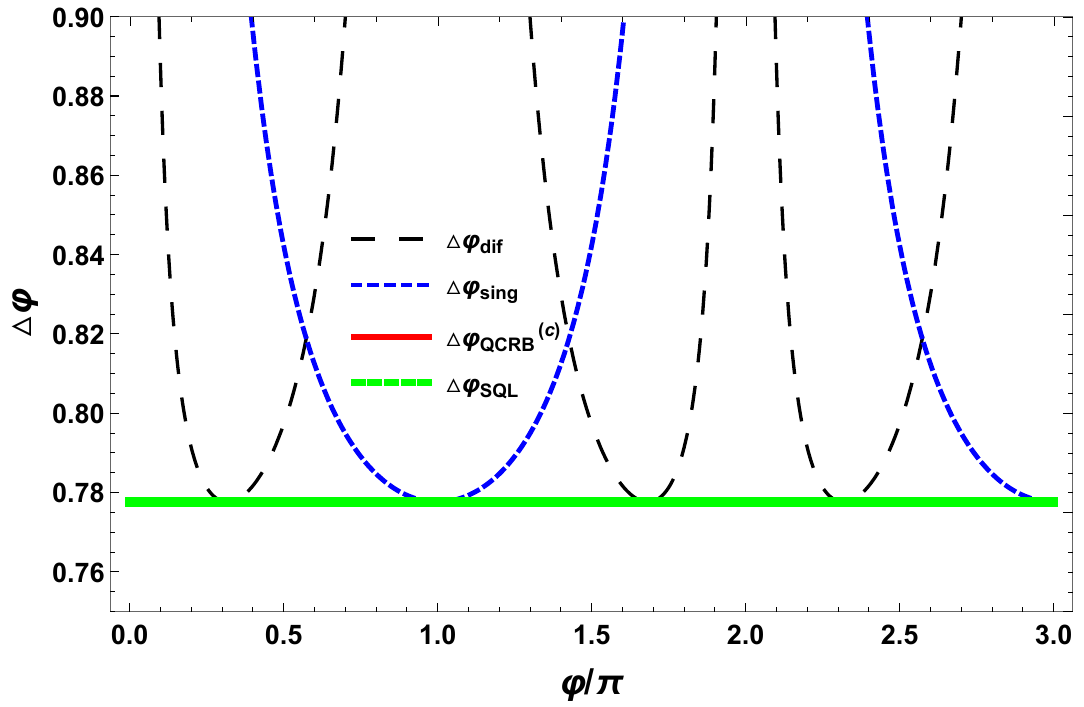} 
  \vfill $\left(b\right)$
	\end{minipage}
	\caption{The phase sensitivity $\Delta\varphi$ concerning the phase shift in the SMI and DI detection schemes, denoted as $\Delta\varphi_{(sing)}$, $\Delta\varphi_{(dif)}$, respectively, together with the QCRB $\Delta\varphi_{QCRB}^{(c)}$ and the SQL $\Delta\varphi_{SQL}$ with $j=1$. In (a), the parameters $\tau =0.9$ and $\tau '=0.1$. In (b), $\tau=1/\sqrt{2}$ and $r=i/\sqrt{2}$. The graphical representation includes a dashed black line for $\Delta\varphi_{(dif)}$, a long dashed blue line for $\Delta\varphi_{(sing)}$, an orange line for $\Delta\varphi_{QCRB}^{(c)}$, and a green line for $\Delta\varphi_{SQL}$.} \label{The variation of phase sensitivity and BCRB}
	\end{figure}
 
\begin{figure}[ht] 
	\begin{minipage}[b]{.50\linewidth}
		\centering
		
		\includegraphics[scale=0.419]{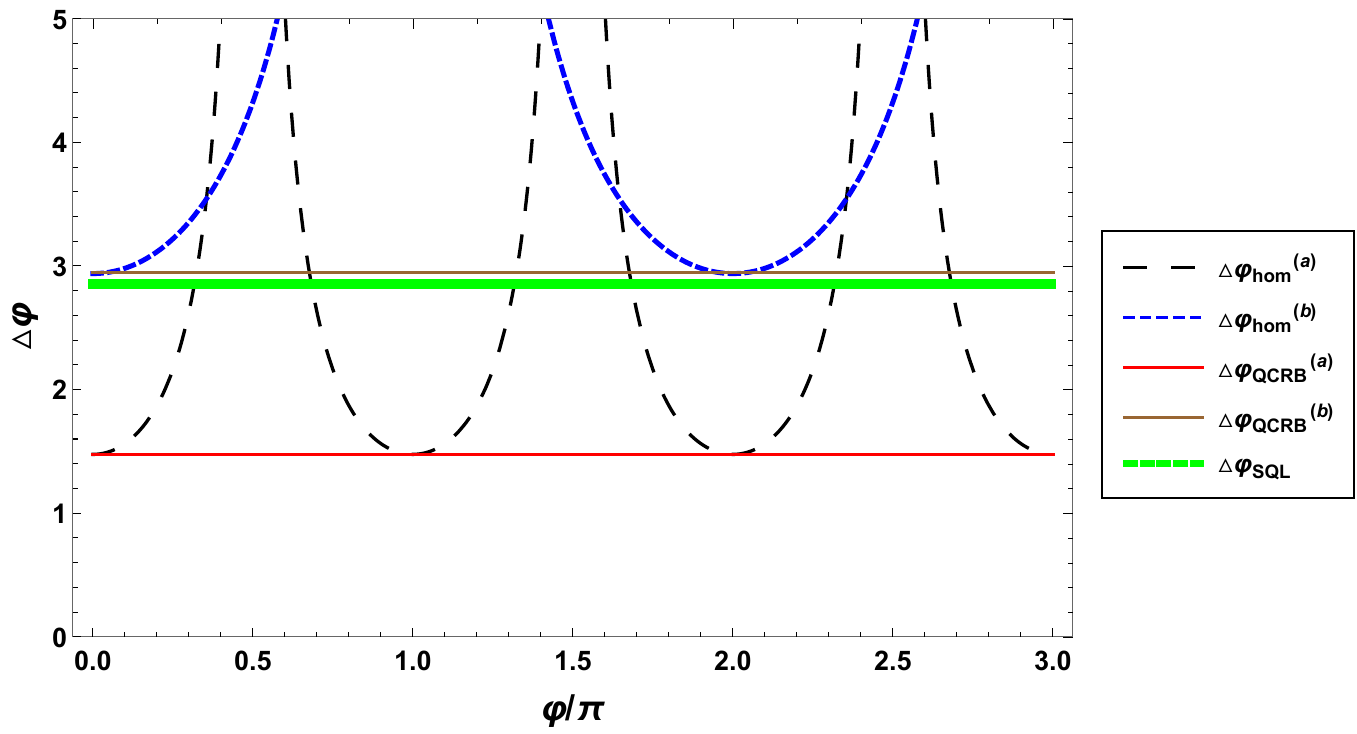} 
  \vfill $\left(a\right)$
	\end{minipage} \quad
	\begin{minipage}[b]{.45\linewidth}
		\centering
		\includegraphics[scale=0.419]{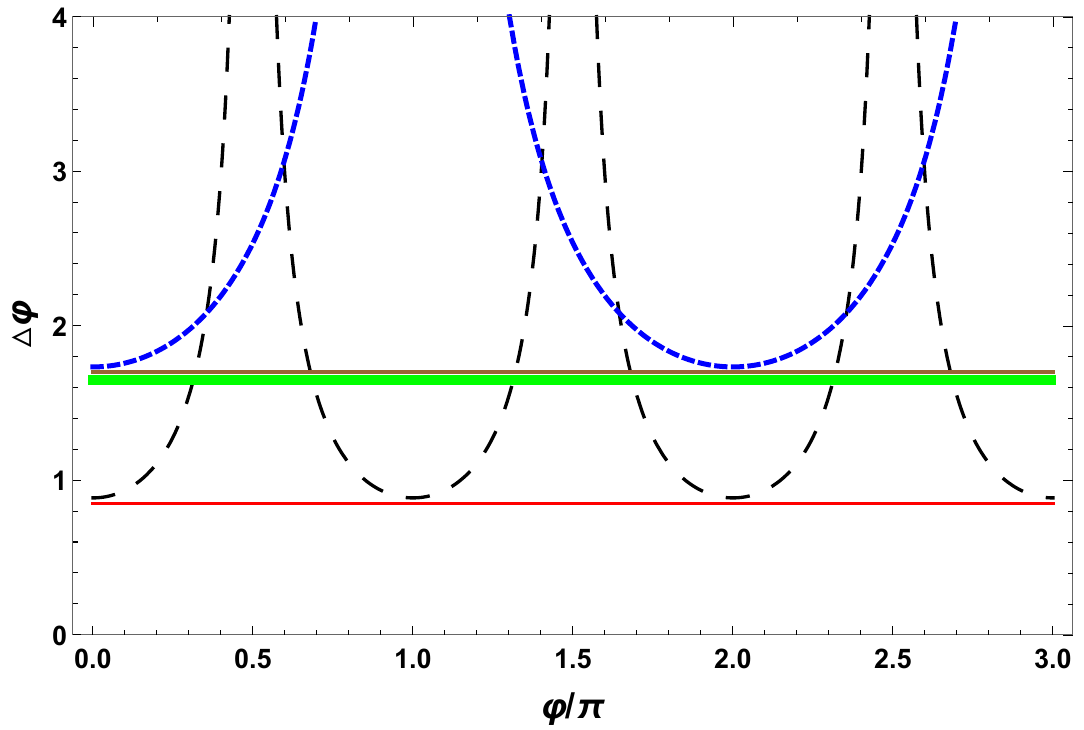} 
  \vfill $\left(b\right)$
	\end{minipage}
	\caption{The phase sensitivity $\Delta\varphi$ concerning the phase shift in the BH detection schemes, denoted as $\Delta\varphi_{(Hom)}$, together with the QCRB $\Delta\varphi_{QCRB}^{(a)}$ and $\Delta\varphi_{QCRB}^{(b)}$ discussed previously with $j=1$ in (a) and $j=3$ in (b), upon the fixed parameters $\tau =0.99$ and $\tau '=0.01$. The graphical representation includes a long dashed black line for $\Delta\varphi_{Hom}^{(a)}$, a long dashed blue line for $\Delta\varphi_{Hom}^{(b)}$, a red line for $\Delta\varphi_{QCRB}^{(a)}$, a brown line for $\Delta\varphi_{QCRB}^{(b)}$ and a green line for the SQL.} \label{Fig4}
	\end{figure}	
 
\end{widetext}

On the other hand, FIG \ref{The variation of phase sensitivity and BCRB}(b), we illustrate the case of a balanced beam splitter, i.e., $\tau=1/\sqrt{2}$ and $r=i/\sqrt{2}$, where $j=1$ considering its optimal setup.
The blue and black curves correspond to the sensitivity variation in the SMI and DI detection schemes, respectively. The solid orange lines are for $\Delta_{QCRB}^{(c)}$, and the solid green lines represent the SQL. This figure shows that curves have an optimum that achieves the QCRB and attends the SQL, which the two-parameter QFI implies.

For the BH detection scheme, we consider transmission coefficients $\tau=1$ for the first beam splitter and $\tau'=0$ for the second beam splitter. Then, the two graphs indicted in FIG \ref{Fig4} (a) and (b) unveil intriguing patterns. Both share a common x-axis, labeled $\varphi/\pi$, from $0$ to $3$. On the y-axis, labeled $\Delta\varphi$. Graph \ref{Fig4} (a) shows cases $\Delta_{QCRB}$ and $\Delta\varphi_{Hom}$, represented by dashed and dotted lines, respectively when $j=1$. Their symmetrical oscillations peak near $4$ for $\Delta\varphi$, hinting at phase uncertainty dynamics. Furthermore, the solid green line, $\Delta_{QCRBs}$, remains steadfastly above 2, perhaps denoting a theoretical limit. Moreover, SQL is introduced, which serves as a benchmark for comparison. Graph \ref{Fig4} (b) echoes these themes but with subtler oscillations and a lower $\Delta_{QCRBs}$ value for $j=3$. This result shows that the BH detection scheme approaches the QCRBs corresponding to $\mathcal{F}^{(a)}$ and $\mathcal{F}^{(b)}$. All QCRB limits have a physical meaning and can actually be attained with the appropriate setup. The introduction of the SQL provides a valuable reference point, emphasizing the performance of the BH detection scheme in relation to fundamental quantum limits. \par

To sum up, Fig.\ref{The variation of phase sensitivity and BCRB} and \ref{Fig4} provide beneficial visuals for understanding how detection schemes work in quantum interferometry. They don't just show how close these schemes get to the theoretical limits set by the QCRBs; they also help design and improve experiments. By showing us the complex relationship between detection methods and input states, these figures help us understand how to improve experimental outcomes and advance technology. Our results align with previous findings that underscore the importance of external phase control. Specifically, we observed that the external phase adjustments significantly enhance the measurement precision, corroborating the theoretical predictions and experimental outcomes reported by Zhang et al \cite{Stefan}. and others in the field \cite{ Zhang 2020, Adhikari 2019}. These findings suggest that meticulous management of external phase parameters is integral to advancing the accuracy of quantum measurement protocols.\par
 
 \section{Concluding Remarks and Outlook}\label{sec7}
In conclusion, this paper presents theoretical computations of the QFI in both scenarios involving two parameters and those with a single parameter. We employed coherent input states within the SU(2) Lie algebra framework. This study delved into the fundamental limits of precision in quantum parameter estimation governed by QCRB and the QFI. We focused on interferometric phase sensitivity, a crucial aspect of quantum measurement techniques. By employing high-intensity lasers within a MZI, we aimed to enhance the accuracy of phase measurements. Through a comparative analysis between single-parameter and two-parameter QFIs, we explored the phase sensitivity of an MZI under three distinct detection schemes: single-mode intensity detection, difference intensity detection, and balanced homodyne detection. Throughout our investigation, we utilized a spin-coherent state associated with the su(2) algebra as the input state across all scenarios.\par

Notably, our findings reveal that all three detection schemes, when coupled with the spin-coherent input state, can attain the QCRB. This underscores the efficacy of these detection methodologies in achieving optimal precision in phase estimation within quantum systems. Additionally, our result emphasizes that an external phase reference is critical in quantum interferometry, significantly enhancing phase estimation precision and impacting both theoretical and practical outcomes. Eventually, our study contributes valuable insights into quantum parameter estimation, shedding light on the interplay between detection schemes, input states, and achievable precision limits. Such understanding is crucial for advancing quantum measurement techniques and realizing their full potential in various scientific and technological domains.

\end{document}